\documentclass[11pt]{article}

\usepackage[T1]{fontenc}
\usepackage[utf8]{inputenc}
\usepackage{geometry}
\usepackage{amsmath,amssymb}
\usepackage{graphicx}
\usepackage{booktabs}
\usepackage{comment}
\usepackage{algorithm}
\usepackage{algpseudocode}
\usepackage{caption}
\usepackage[most]{tcolorbox}
\usepackage{url}
\usepackage[hidelinks]{hyperref}

\title{Information-theoretic formulation of the Traveling Salesman Problem}

\author{
	Enrico Maria Fenoaltea$^{1,2}$, Riccardo Piombo$^{1}$, Aurelio Patelli$^{1}$\\[0.6em]
	\small $^{1}$Centro Ricerche Enrico Fermi (CREF), Via Panisperna 89 A -- 00184 Roma, Italy\\
	\small $^{2}$Universitat de Barcelona Institute of Complex Systems (UBICS),\\
	\small Universitat de Barcelona, 08028 Barcelona, Spain
}

\date{\today}

\begin{document}
	
	\maketitle
	
	\begin{abstract}
		The Traveling Salesman Problem (TSP) asks for the shortest route to visit a set of cities exactly once. It combines a simple local rule - each city must be visited once - with a hard, global constraint- all cities must be traversed within a single cycle. 
		We cast the problem within a probabilistic, information-theoretic framework. The coexistence of local and global constraints is precisely what makes the problem difficult to address in this framework: the local rule can be enforced through vertex-level constraints, whereas the global constraint cannot be captured by independent edge probabilities. We show that this obstacle can be overcome by defining a maximum-entropy probability distribution over graphs, in which edge costs and degree constraints generate an assignment-like ensemble, and a global term, describing the hard constraint, tilts this ensemble toward Hamiltonian cycles. 
		To make the construction tractable, we derive a mean-field approximation in terms of edge occupancies and implement a differentiable cycle penalty that suppresses sub-tours. 
		This leads to a self-consistent numerical procedure whose output is not only a candidate tour but also a probability matrix encoding competing edges and degenerate solutions. We test the method on synthetic ensembles 
		and on TSPLIB instances. The algorithm converges to connected tours in polynomial time, matching the best-known solution in the majority of instances and remaining within a small relative gap otherwise. Beyond its competitive performance, the proposed framework offers a general approach for handling hard constraints while reducing hard combinatorial optimization problems to simpler ones.
	\end{abstract}

	\noindent\textbf{Keywords:} TSP; Random Graphs; Information Theory; Statistical Physics
	
	\section{Introduction}
	
	The Traveling Salesman Problem (TSP) is one of the simplest optimization problems to formulate and one of the most influential in the theory of computation~\cite{karp2009reducibility,garey2002computers,cook2011traveling}, graph theory~\cite{bondy1976graph}, and statistical physics~\cite{vannimenus1984statistical,mezard1985replicas}.
	Given a set of points and pairwise travel costs, such as cities connected by routes, the goal is to find a minimum‑cost closed tour visiting each point exactly once.
	The apparent simplicity of this task conceals two distinct conditions. 
	A valid tour is characterized simultaneously by a local structure and a global pattern: locally, each vertex must be incident to exactly two edges; globally, all vertices must be connected in a single cycle, known as a Hamiltonian cycle~\cite{bondy1976graph,baniasadi2018new}.
	The difficulty of the TSP arises from minimizing the total cost while enforcing both local and global constraints simultaneously.
	
	Solutions to the TSP are typically sought either directly within the discrete space of feasible tours or through relaxations of the corresponding feasible set. Integer-programming formulations, for example, combine degree constraints with mechanisms for eliminating subtours~\cite{miller1960integer}. Geometric and polyhedral approaches investigate the structure of the feasible set and its convex hull, whereas algorithmic approaches include exact methods, approximation algorithms, heuristics, mathematical relaxations~\cite{pan2023h}, and machine-learning-based methods~\cite{parjadis2023learning}.

	Together, these approaches constitute a rich body of theory and methodology for the TSP. Yet they also tend to treat the TSP as a rigid hard constraint-satisfaction problem over discrete structures. For a detailed overview of the principal approaches to the problem, see Ref.~\cite{applegate2011traveling}. 
	
	Another important class of methods comprises metaheuristics, including genetic algorithms \cite{potvin1996genetic, razali2011genetic} and ant-colony optimization \cite{dorigo1997ant}, as well as statistical-physics-inspired approaches that are closer in spirit to the framework developed below, such as simulated annealing \cite{kirkpatrick1983optimization, hopfield1985neural}, mean-field annealing \cite{bilbro1988optimization}, and softassign \cite{gold1996softmax, gold1995softassign}. 
	These methods likewise search over tours or permutations. In classical statistical-physics formulations, this search is typically expressed through a position-based encoding of the tour, which leads to an Ising-like Hamiltonian closely related to a quadratic assignment problem \cite{loiola2007survey}.
	
	We instead introduce a (complementary) probabilistic view of the TSP,  formulating it as an inference problem over a probability distribution defined on edge-weighted graphs.
	Rather than searching directly for the optimal tour, we ask which probability distribution over graphs best represents the same local and global structures mentioned above, without introducing further unwarranted assumptions. 
	This reframes the TSP as an inference task, formulated in the language of random graph theory and information theory, which naturally leads to a variational formulation of maximum-entropy~\cite{cimini2019statistical}:
	the local degree constraints enter as expectation constraints on vertex degrees, while the objective cost enters as an observable similar to energy~\cite{buffa2025maximum}. 
	In the absence of a global constraints forcing an Hamiltonian cycle, the resulting distribution has the edge-factorized form of the probabilistic formulations of matching and assignment problems~\cite{baybusinov2026grand}. 
	The remaining difficulty is precisely the one that distinguishes the TSP from those problems: a Hamiltonian tour is not merely a set of locally compatible edges, but a single connected component spanning all vertices of the graph.
	
	We address this global requirement by introducing, within this ensemble, a penalty that favors connected configurations. 
	This penalty increases the statistical weight of graphs with the correct global structure and suppresses those that satisfy the local degree constraints but decompose into disconnected subtours; in an appropriate limiting regime, this suppression becomes complete, and only connected graphs survive. Since a two-regular connected graph on all vertices is a Hamiltonian cycle, the TSP can be recovered as a zero-temperature and strong-penalty limit of this probability distribution over graphs. The central challenge is that the local terms, being additive over edges, naturally yield a factorized probability law, whereas the global penalty depends on the graph as a whole - so the mathematical task is to construct a distribution that keeps this local factorization explicit while isolating the global term. We now formulate this construction using the maximum-entropy principle.
	
	\section{Information theory framework}
	We tackle the TSP by defining a  statistical space, in which every graph is assigned a probability, and the tour emerges only as a limiting configuration.
	Let $G$ denote a graph of $n$ vertices, with adjacency matrix $G_{ij}\in\{0,1\}$ and edge costs $c_{ij}$.
	The variational formulation follows from the maximum-entropy principle applied to graphs~\cite{newman2018networks,saracco2015randomizing,cimini2019statistical}.
	We seek the probability distribution $P(G)$ over graph configurations that is maximally noncommittal while satisfying the constraints imposed by the problem: normalization, energetic cost, local degree constraints, and a global penalty enforcing a unique connected cycle.
	For the general case of a directed graph, the degree constraints, imposing one incoming and one outgoing edge, are written as
	\begin{equation}
		\label{eq:local_constraints}
		\begin{aligned}
			\sum_{j\neq i}\langle G_{ij}\rangle_P &= 1,
			\qquad i=1,\ldots,n,\\
			\sum_{i\neq j}\langle G_{ij}\rangle_P &= 1,
			\qquad j=1,\ldots,n.
		\end{aligned}
	\end{equation}
	where $\langle \cdot \rangle_P = \sum_G (\cdot) P(G)$ denotes the average over the probability distribution $P$.
	The global constraint is represented by a graph functional $N_\text{conn}(G)$, whose associated coupling parameter is denoted by $\mu$.
	At this stage, its detailed form does not need to be specified.
	It is sufficient to assume that, in an appropriate limit, this term, when imposed together with the constraints in Eq.(\ref{eq:local_constraints}), favors graphs formed by a unique connected cycle.
	
	The Lagrangian associated to the problem is
	\begin{equation}
		\label{eq:lagrangian_probability}
		\begin{aligned}
			\mathcal{L}_\mu[P] ={}&
			-\langle\log P(G)\rangle_P
			-\beta\left\langle\sum_{i,j}G_{ij}c_{ij}\right\rangle_P\\
			&+\alpha\langle 1\rangle_P
			+\sum_i u_i\left\langle 1-\sum_{j\neq i}G_{ij}\right\rangle_P\\
			&+\sum_j v_j\left\langle 1-\sum_{i\neq j}G_{ij}\right\rangle_P
			-\mu\langle N_{\mathrm{conn}}(G)\rangle_P .
		\end{aligned}
	\end{equation}

	The terms in Eq.~(\ref{eq:lagrangian_probability}) have a direct interpretation.
	The first term is the Shannon entropy.
	The second term is the energetic cost of the graph.
	The multiplier $\alpha$ enforces probability density normalization.
	The multipliers $u_i$ and $v_j$ impose the local degree constraints.
	Finally, the last term introduces the global bias that assigns larger weight to graphs with the required single cycle structure. The advantage of this formulation is that, at finite $\beta$, the entropy term turns the hard discrete optimization problem into a probabilistic problem. This yields an explicit Gibbs ensemble over graphs. The resulting formulation is therefore more tractable analytically and more advantageous algorithmically, as we shall show below, while the original deterministic problem is formally recovered in the zero-temperature limit $\beta\to\infty$, where the entropic contribution becomes negligible.
	
	\begin{figure}[!t]
		\centering75	\includegraphics[width=0.8\linewidth]{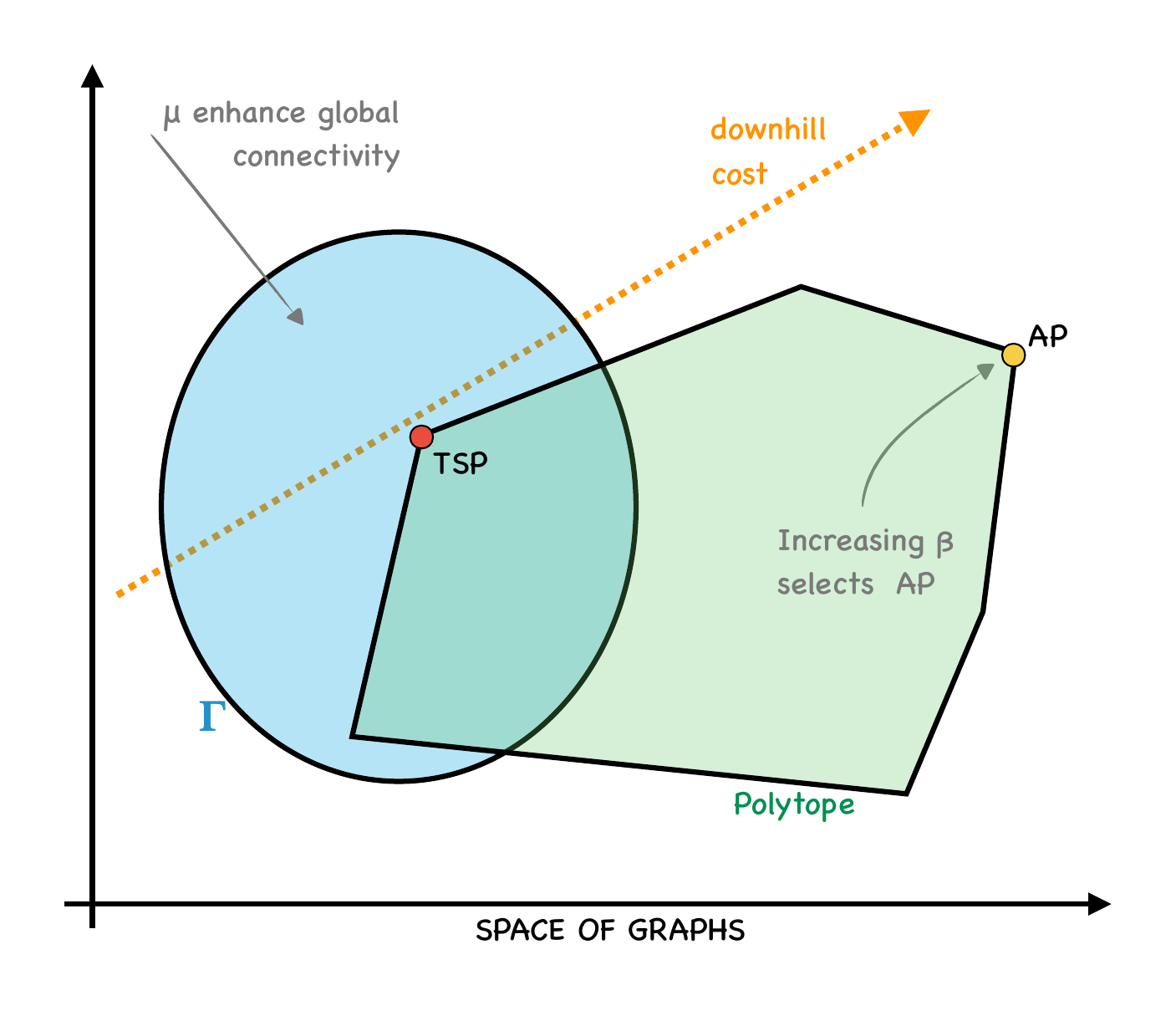}
		\caption{Schematic representation of the space of graphs considered in the probabilistic formulation. The assignment polytope represents the region in which the local degree constraints are satisfied. Within this region, the assignment-problem solution, denoted by \textbf{AP}, is selected by increasing the inverse temperature $\beta$ when only local constraints and edge costs are considered. The TSP solution, denoted by \textbf{TSP}, belongs to a region that is an intersection between the AP polytope and the $\Gamma$ region. Indeed $\Gamma$ is defined as a region that, when the degree constraints in Eq.\ref{eq:local_constraints} are satisfied, contains only hamiltonian cycles. Varying $\mu$ reinforces the probability of this globally region. The figure is intended as a conceptual guide rather than as a faithful geometric representation of the graph space.} 
		\label{fig:space-graph}
	\end{figure}
	Figure~\ref{fig:space-graph} summarizes the geometric intuition behind the construction.
	The local degree constraints define a feasible polytope of sparse graph configurations.
	If only these local constraints and the edge costs are considered, increasing the relevance of the cost, parametrized by $\beta$, drives the probability distribution toward a vertex of the polytope or an Assignment-Problem solution (marked by AP in the figure).
	However, this solution need not be a Hamiltonian cycle: it may consist of several disconnected subtours.
	The additional role of the global term, parametrized by $\mu$, is to tilt the probability distribution toward the region of Hamiltonian cycle graphs.
	
	Maximizing Eq.~(\ref{eq:lagrangian_probability}) with respect to $P(G)$ gives a probability distribution on graph space,
	\begin{equation}
		P_\mu^\star(G) = e^{-\mu  N_\text{conn}(G) - \log\langle e^{-\mu N_\text{conn}(G)}\rangle_{P_0}} P_0(G),
		\label{eq:probability_tsp}
	\end{equation}
	where we have isolated the factorized contribution
	\begin{eqnarray}
		\label{eq:probability_ap}
		P_0(G)&=&  \prod_{i\neq j}p_{ij}^{G_{ij}}\left(1-p_{ij}\right)^{1-G_{ij}}\\
		p_{ij}&=&\frac{e^{-\beta c_{ij} + u_i + v_j}}{1+e^{-\beta c_{ij} + u_i + v_j}} ,\nonumber
	\end{eqnarray}
	which corresponds to the maximally noncommittal probability distribution of the assignment-like problem.
	The multipliers $\left\{u_i, v_j\right\}$ are determined by solving the degree constraints in Eq.~(\ref{eq:local_constraints}).
	
	Equations~(\ref{eq:probability_tsp}) and~(\ref{eq:probability_ap}) make explicit the separation between the assignment-like structure and the global TSP correction.
	The factorized probability $P_0(G)$ contains the local information: edge costs and degree constraints. Taken alone, this term defines the probabilistic solution of the assignment problem: $p_{ij}$ is the probability that the edge $i\to j$ is selected, and in the zero-temperature limit $\beta\to\infty$, the probability matrix becomes binary, selecting the optimal assignment \cite{baybusinov2026grand}.
	
	The remaining factor,
	\begin{equation}
		e^{-\mu  N_\text{conn}(G) - \log\langle e^{-\mu N_\text{conn}(G)}\rangle_{P_0}},
	\end{equation}
	is the exponential tilt of the above assignment-like ensemble.
	Its role is to reshape the probability distribution by increasing the statistical weight of graphs with the required topology.
	Thus, the TSP is represented as an assignment-like probability distribution coupled to a graph-level correction.
	The limiting case of interest is the one in which this becomes a projector  $\chi_\Gamma(G)$ onto the region $\Gamma$ of graphs that, when the local degree constraints are satisfied, are Hamiltonian cycles.
	More precisely, we seek a limit of $\mu$ such that
	\begin{equation}
		P_\mu^\star(G)
		\longrightarrow
		\chi_\Gamma(G)P_0(G).
	\end{equation}
	In this regime, the distribution is no longer the unconstrained assignment-like ensemble, but the same ensemble restricted to the unique cycles sector.
	Then, the zero-temperature limit, obtained by increasing $\beta$, selects the minimum-cost tour.
	
	The limit that realizes this projection depends on the behavior of $N_\text{conn}(G)$.
	In the following, we assume that it is a non-negative penalty that, when the degree constraints are satisfied, vanishes inside the Hamiltonian cycles region $\Gamma$ \footnote{A second class of connectivity terms considers the case in which $N_\text{conn}(G)$ is zero outside $\Gamma$ and positive inside. This alternative construction will be discussed in the Supplementary Material, since it is not used in the numerical implementations.},
	\begin{equation}
		N_\text{conn}(G)\begin{cases}
			=0 &\text{for  } G\in\Gamma\\
			>0 &\text{otherwise}
		\end{cases}
	\end{equation}
	so that $\Gamma$ is selected by taking $\mu\to\infty$.
	Before this limiting regime is reached, $\mu$ controls the strength with which the assignment-like ensemble is tilted toward the correct topology.
	
	The maximum-entropy construction therefore provides a formal probabilistic solution to the separation between local and global components, together with a clear interpretation. At the same time, although $P_0(G)$ factorizes over edges, the tilting factor depends on the graph as a whole. Consequently, the local constraints become coupled through a nonlocal connectivity term. The formal construction thus converts the original combinatorial constraint into a statistical-mechanical one: the probability law is explicit, but its direct evaluation remains difficult. This is the point at which an approximation becomes necessary to make the tilted ensemble computationally accessible while preserving its assignment-like structure as much as possible.

	\section{Mean-field approximation}
	A consistent approximation of Eq.~(\ref{eq:probability_tsp}) shall preserve the factorized structure of the local ensemble while retaining, in an effective form, the influence of the global penalty.
	The mean-field approximation implements this idea by restricting the exact variational principle to a simpler family of trial measures. In particular, the problem is factorized by replacing $P(G)$ with
	\begin{equation}
		Q(G)=\prod_{i\neq j} q_{ij}(G_{ij}) = \prod_{i\neq j}
		V_{ij}^{\,G_{ij}}
		(1-V_{ij})^{1-G_{ij}},
		\label{eq:factorized_trial_measure}
	\end{equation}
	considering the mean adjacency matrix
	\begin{equation}
		V_{ij}=\langle G_{ij}\rangle_Q,\quad  \Rightarrow \quad  0\le V_{ij}\le 1.
	\end{equation}
	The entries $V_{ij}$ are therefore edge occupancy probabilities and, since $G_{ij}$ is binary, we can write $V_{ij}=Q(G_{ij}=1)$.
	Under the mean-field ansatz, correlations between distinct edges are neglected, i.e., $\langle G_{e_1}G_{e_2}\rangle_Q \approx \langle G_{e_1}\rangle_Q \langle G_{e_2}\rangle_Q$.
	
	The mean-field variational problem is obtained by restricting the original problem to the manifold of factorized measures,
	\begin{equation}
		\mathcal{L}_{\mathrm{MF}}(V)
		:=
		\mathcal{L}_\mu[Q_V].
		\label{eq:mf_functional_definition}
	\end{equation}
	
	In this way, the original optimization over all graph distributions $P(G)$ is replaced by a simpler finite-dimensional optimization over the edge marginals $V_{ij}$.
	This construction is directly analogous to standard mean-field theory in statistical mechanics~\cite{cardy1996scaling}.
	Maximizing this Lagrangian function with respect to $V_{ij}$ gives the self-consistency equations
	\begin{equation}
		V_{ij} = \frac{1}{ 1+e^{\beta c_{ij}-u_i-v_j+\mu\tfrac{\partial N_\text{conn}(V)}{\partial V_{ij}}} }.
		\label{eq:mf_self_consistency}
	\end{equation}
	Together with the local constraints of Eq.~(\ref{eq:local_constraints}), these equations define the mean-field problem.
	The edge probability $V_{ij}$ is controlled by three contributions: the cost of the edge tuned by the inverse temperature $\beta$, the degree-enforcing fields represented by the Lagrange multipliers $u$ and $v$, and the effective global field generated by the penalty term and tuned by $\mu$.
	
	When $\mu=0$, Eq.~(\ref{eq:mf_self_consistency}) reduces to the assignment-like probability of Eq.~(\ref{eq:probability_ap}).
	For $\mu>0$, instead, the last term in the exponent penalizes the $V_{ij}$s that increase the penalty.
	
	In this way, the global constraint encoded by $N_{\mathrm{conn}}(V)$ 
	now plays the role of an effective field acting locally on individual edges. Hence, the above equation retains the edge-wise structure of the linear assignment ensemble, while each edge is still affected by the global topology of the mean graph.
	
	The mean-field approximation neglects these loop-induced correlations and retains only the contribution of the first moments.
	This is the only step at which the global nature of the TSP constraint is approximated.
	The local assignment-like part is already factorized by the maximum-entropy construction, and the heterogeneity induced by the degree constraints is already absorbed into the Lagrange multipliers $u$ and $v$.
	
	Although there is no universal small parameter controlling the mean-field approximation, the approximation is not required to reproduce the entire finite-$\mu$ probability distribution accurately. The global penalty term is intended to impose a hard constraint in the limit $\mu\to\infty$. In this limit, what matters primarily is whether the penalty vanishes, rather than its precise value for invalid configurations. The essential requirement is therefore that the mean-field surrogate have the same zero-penalty sector as the original term, namely the sector corresponding to Hamiltonian cycles in the relaxed space. If this condition is satisfied, the approximation may still correctly describe the limiting constrained problem even if it does not accurately reproduce the probabilities of all graphs at finite $\mu$.

	Now, A natural question is whether the mean-field closure is not only formally consistent, but also practically useful as a search principle. In practice, given a cost matrix $C_{ij}$, our goal is to solve Eq.\ref{eq:mf_self_consistency} for the edge-occupation variables $V_{ij}$ at sufficiently large values of $\beta$, to favor low-cost configurations, and $\mu$, to enforce a single valid cycle. An optimal tour can then be extracted from the resulting probabilistic solution. To this end, in the following, we define a differentiable global penalty term $N_{\mathrm{conn}}(V)$ and benchmark the tours obtained from the self-consistent equations on both synthetic and real TSP cost matrices.
	
	\section{Numerical study in the mean-field approximation}
	Turning the above mean-field variational approach into a numerical procedure requires a concrete and differentiable choice of the global penalty term $N_{\mathrm{conn}}(V)$. As discussed above, in the absence of this global term, the $\beta\to\infty$ solution of Eq.~\ref{eq:mf_self_consistency} is a permutation matrix which in general decomposes into a collection of disjoint directed cycles. A valid TSP solution corresponds to the special case in which this decomposition contains a single cycle of length $N$. This condition is equivalent to requiring all proper subgraphs of the assignment to be directed acyclic graphs (DAGs) \cite{zhang2022truncated, zhang2025analytic, zhang2026solving}. Therefore, the role of $N_{\mathrm{conn}}(V)$ is to suppress all cycles of length smaller than $N$, while leaving the Hamiltonian cycle unconstrained. Since the trace of the $l$-th power of the adjacency matrix, $\operatorname{Tr}(G^l)$, is positive only when the directed graph contains a closed walk of length $l$, we can write:  
	\begin{equation}
		\langle N_\text{DAG}(G)\rangle_Q= N_{\text{DAG}}(V) = \sum_{l=2}^{n-1} \alpha_l \frac{\operatorname{Tr}\left(V^l\right)}{l},
		\label{eq:dag_penalty}
	\end{equation}
	where $\alpha_l\geq 0$ controls the strength of the penalty associated with the presence of cycles of length $l$.
	Hence, in the strong-penalty and low-temperature regime, where both $\mu$ and $\beta$ are large, the mean graph is expected to concentrate around a low-cost Hamiltonian tour.
	
	The numerical procedure alternates between evaluating the effective subtour-elimination field generated by the current matrix $V$ and solving the corresponding mean-field equations in Eq.~(\ref{eq:mf_self_consistency}). Starting from an arbitrary initial condition $V^0$, the process is iterated until convergence, defined as the point at which the matrix $V$ no longer changes appreciably. The worst-case computational cost of this procedure is estimated to scale as $N^4$. The complete pseudo-code with more details on the numerical implementation are reported in the Methods section.
	Unless otherwise stated, we use positive and uniform weights $\alpha_l$ for all cycle lengths in Eq.~(\ref{eq:dag_penalty}).
	
	An important consequence of this procedure is that its output is not limited to the single cycle eventually extracted from the solution. Indeed, the weighted adjacency matrix $V$, being a probability matrix, retains information about competing edges and alternative solutions.
	This additional information allows us to examine separately two structural features of the cost matrix that can make the search problem difficult: the degeneracy of the optimum and correlations among edge costs.
	Although we isolate them in controlled numerical experiments, these two features are not mutually exclusive and may occur simultaneously in structured or application-driven instances.
	
	\begin{figure}[t]
		\centering
		\includegraphics[width=0.75\linewidth]{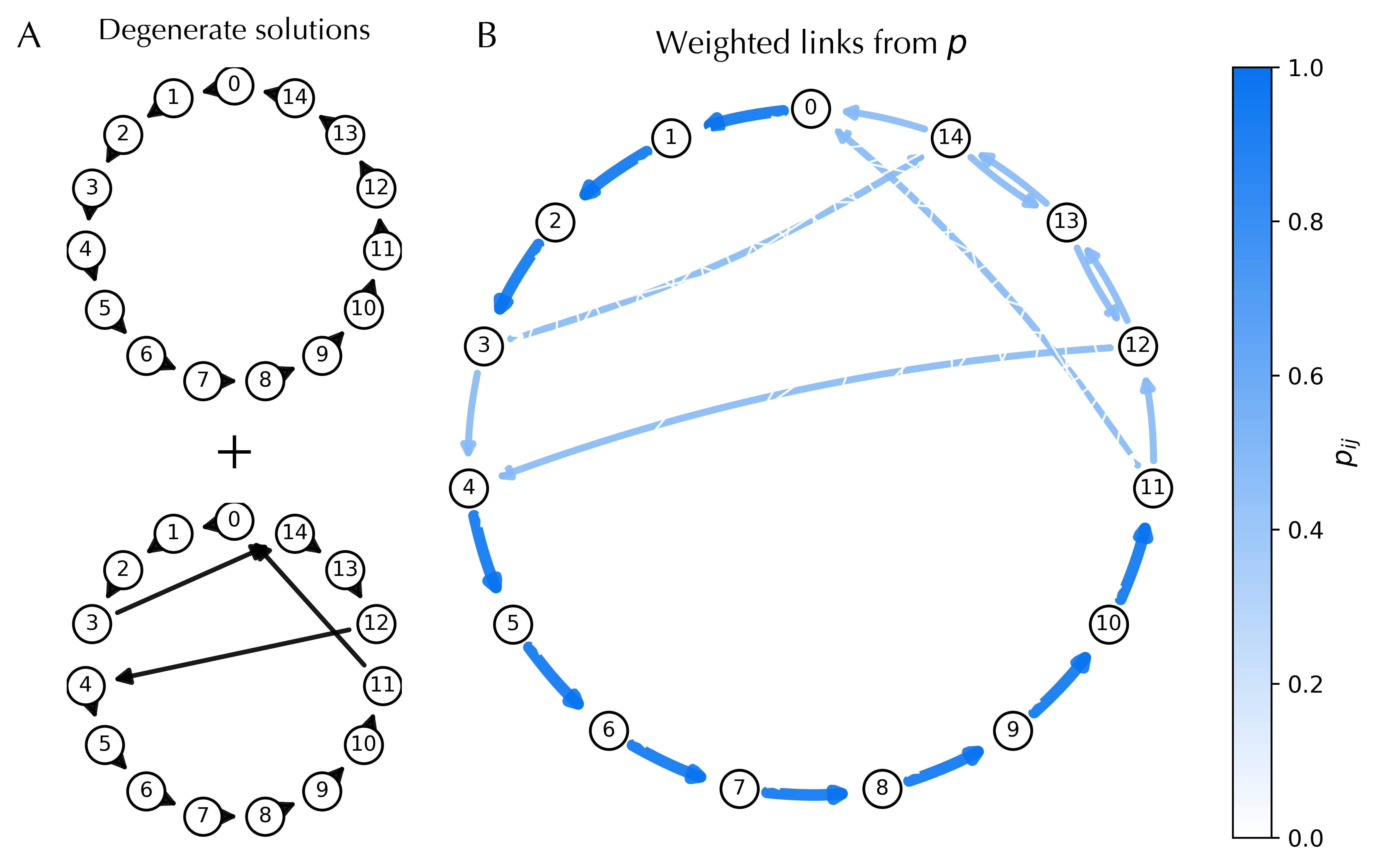}
		\caption{(A) Two equal-cost Hamiltonian cycles used to construct the degenerate cost matrix, displayed on a circular vertex layout ordered according to the first cycle. (B) Corresponding mean-field matrix $p_{ij}$ on the same layout; link color scale and thickness indicate edge probability.}
		\label{fig:degenerate}
	\end{figure}
	
	Degeneracy is the most direct setting in which the probabilistic character of the formulation becomes relevant. For a non-degenerate cost matrix, where the optimal TSP solution is unique, the converged matrix $V$ becomes binary in the zero-temperature limit $\beta\to\infty$, as in the assignment problem~\cite{baybusinov2026grand,koehl2021fast}. In this case, the probability that an edge belongs to the optimal solution is either zero or one. By contrast, when several Hamiltonian cycles have exactly the same minimum cost, the converged matrix $V$ could be not-binary even in the low temperature limit. Instead, its entries $V_{ij}$ can be interpreted as the probability that edge $i\to j$ appears in an optimal tour chosen from the degenerate set. Thus, while a deterministic algorithm must ultimately return a single representative of the degenerate solutions, the mean-field formulation can retain information about several equivalent optima.
	
	To illustrate this, we construct an ad hoc cost-matrix instance with two optimal TSP solutions, corresponding to two Hamiltonian cycles with exactly the same minimum cost.
	The two solutions are shown in Fig.~\ref{fig:degenerate}A using a common circular arrangement of the vertices.
	As shown in Fig.~\ref{fig:degenerate}B, when the initialization $V^0$ does not favor either solution (e.g. $V^0_{ij}=1/N$ for any $i,j$), the converged mean-field matrix does not collapse onto a binary permutation corresponding to a single cycle.
	Instead, it assigns larger probabilities to the edges belonging to the two optimal tours, with their relative weights reflecting how frequently the edges occur within the degenerate optimal set.
	The output is therefore a weighted superposition of equivalent solutions rather than an arbitrary selection among them.
	
	\begin{table*}[!t]
		\centering
		\caption{Comparison of relative cost gaps across competing algorithms for random instance.}
		\begin{tabular*}{\textwidth}{@{\extracolsep{\fill}} |ll|cc|cc|cc|cc| }
			\toprule 
			N & algorithm & \multicolumn{2}{c}{random asymmetric} & \multicolumn{2}{c}{correlated asymmetric} & \multicolumn{2}{c}{random symmetric} & \multicolumn{2}{r}{correlated symmetric} \\ 
			&  & median & counting & median & counting & median & counting & median & counting \\
			\midrule
			20 & TGCA & 0.00\% & 11:68:1 & 0.00\% & 49:49:2 & 35.20\% & 70:2:0 & no data & no data \\
			& LKH & 0.00\% & 0:75:5 & 0.00\% & 5:86:9 & 0.00\% & 1:93:6 & 0.00\% & 4:86:10 \\
			\cline{1-10}
			50 & TGCA & 0.21\% & 52:37:2 & 0.06\% & 67:27:6 & no data & no data & no data & no data \\
			& LKH & 0.00\% & 0:77:14 & 0.00\% & 5:56:39 & -0.32\% & 1:31:68 & -0.03\% & 1:19:80 \\
			\cline{1-10}
			100 & TGCA & 1.84\% & 74:14:2 & 0.03\% & 70:11:18 & no data & no data & no data & no data \\
			& LKH & 0.00\% & 0:55:35 & -0.01\% & 1:28:70 & -1.24\% & 0:1:99 & -0.07\% & 0:0:100 \\
			\cline{1-10}
			200 & TGCA & 0.39\% & 65:6:27 & 0.00\% & 59:3:38 & no data & no data & no data & no data \\
			& LKH & -0.11\% & 0:26:72 & -0.01\% & 0:7:93 & -1.54\% & 0:0:100 & -0.08\% & 2:0:97 \\
			\cline{1-10}
		\end{tabular*}
		\label{tab:comparison_random}
		\caption{For each problem size and competing algorithm, the table reports the median relative energy gap and the win--equal--loss counts across the benchmark instances. The three entries in the count column indicate the number of cases, out of 100 test instances, in which our method achieves a lower, equal, or higher best-tour cost than the competing algorithm, respectively. TGCA is implemented directly from the pseudocode reported in \cite{zhang2026solving}, without any additional post-processing. For LKH, we use LKH-3 with 10 independent runs and 1000 trials per run. We report \textit{no data} when the competing algorithm fails to return valid tours without additional post-processing.}
	\end{table*}
	
	Although this probabilistic description retains information that would otherwise be lost in approaches that always select a single tour, and may also provide a criterion to detect degeneracy in a cost-matrix instance, it makes the extraction of a specific tour, and therefore the computation of the TSP optimal cost, more complex. Indeed, when $V$ is far from binary and represents a superposition of several tours, we did not identify a general procedure to disentangle the individual solutions. However, when the initial condition is biased toward one of the degenerate optima, the symmetry is broken. The iterative procedure then converges to a binary matrix $V$ corresponding to that specific solution. 
	Thus, perturbing the initial condition to generate a random bias toward one of the degenerate solutions provides a practical way to extract a representative optimal tour, and therefore to compute the optimal TSP cost in degenerate instances.
	
	A distinct challenge arises when the costs are not independent.
	Independent random matrices provide a useful baseline, but they do not capture the structured relationships among edges that commonly occur when costs are generated by geometry, shared attributes, or other latent constraints.
	Such correlations can favor groups of mutually related edges and may interact non-trivially with the cycle penalty.
	To determine whether our method remains effective in this setting, we compare its performance on independent and correlated random cost matrices.
	We additionally distinguish asymmetric from symmetric costs.
	Symmetry is itself an important structural property: in a symmetric matrix, the two directions of an edge carry the same cost, whereas an asymmetric matrix retains distinct costs for the two orientations.
	Combining these two characteristics leads to four synthetic test ensembles: random asymmetric, correlated asymmetric, random symmetric, and correlated symmetric matrices.
	
	For each available matrix class and problem size, we generate 100 instances and compare the tours returned by our mean-field procedure with those obtained from two reference algorithms in the TSP literature. The first is the Lin--Kernighan--Helsgaun heuristic (LKH), one of the most widely used and effective local-search heuristics for TSP and ATSP instances~\cite{helsgaun2017extension}. The second is the more recent Trace-Guided Cost Augmentation (TGCA) algorithm~\cite{zhang2026solving}, which is a natural comparator for our approach because it also relies on a trace-based DAG criterion to penalize subtours in the asymmetric TSP setting.
	The purpose of this comparison is not to claim a fully optimized solver or to identify the best possible parameter tuning.
	Rather, it is to assess whether the proposed information-theoretic framework already provides a viable search principle, and to identify the classes of instances in which the current implementation succeeds or remains limited.
	
	Table~\ref{tab:comparison_random} reports both the median relative energy gaps and the win:equal:loss counts.
	The comparison with LKH shows that the mean-field method frequently reaches the same solutions especially in the smaller and asymmetric instances, while its median gap generally remains limited even when LKH returns a lower-cost tour. 
	
	Overall, this shows that although our mean-field algorithm has not yet been optimized from a computational or numerical perspective, nor tuned for any specific instance ensemble, it already appears competitive with state-of-the-art methods.
	The comparison with TGCA support this conclusion. Our mean-field method outperforms TGCA across all the considered matrix classes. Moreover, without additional post-processing, TGCA fails to return valid Hamiltonian cycles for most symmetric instances. This indicates that our method handles more effectively the potential difficulties associated with applying a DAG-based criterion to symmetric TSP instances \footnote{In the SI, we provide an alternative formulation of our mean field approach that uses a global penalty term more suitable for symmetric cost instances.}.
	
	In this test, due to computational-time constraints, we do not consider synthetic instances of larger size. However, we evaluate single well-known benchmark matrices with larger $N$, as discussed below. A large-scale analysis of larger instances would require a dedicated optimization of the implementation, the penalty schedule, and the tour-extraction procedure. Instead, the aim of this work is to assess the strengths and limitations of the probabilistic framework itself.
	The synthetic benchmarks should therefore be interpreted as a characterization of the current formulation rather than as a definitive comparison between fully tuned solvers.
	
	Synthetic ensembles allow us to control specific structural features, but they cannot reproduce the full heterogeneity of standard reference instances. To this end, we consider asymmetric instances from TSPLIB~\cite{reinelt1991tsplib}, a standard library of benchmark instances for the TSP and related variants, containing widely used test cases with reported optimal or best-known tour costs.
	We compare the same three algorithms against the best-known cost reported for each problem. 
	
	\begin{table*}[!h]
		\centering
		\caption{Performance comparison on TSPLIB ATSP benchmark instances.}
		\label{tab:atsp_results}
		\begin{tabular}{lrrrrrr}
			\hline
			\textbf{Instance} & \textbf{Dimension} & \textbf{Best-known} & \textbf{Ours} & \textbf{TGCA} & \textbf{LKH (1, 100)} & \textbf{LKH (1,1000)} \\
			\hline
			ftv33  & 34  & 1286  & 1286 & 2142 & 1286 & 1286 \\
			ftv35  & 36  & 1473  & 1473 & 1714 & 1473 & 1473 \\
			ftv38  & 39  & 1530  & 1530 & 2315 & 1530 & 1530 \\
			ftv44  & 45  & 1613  & 1613 & 1822 & 1613 & 1613 \\
			ftv47  & 48  & 1776  & 1776 & 2336 & 1776 & 1776 \\
			ftv55  & 56  & 1608  & 1614 & 2087 & 1608 & 1608 \\
			ftv64  & 65  & 1839  & 1839 & 2572 & 1839 & 1839 \\
			ftv70  & 71  & 1950  & 1950 & 2662 & 1950 & 1950 \\
			ftv90  & 91  & 1579  & 1586 & 3976 & 1579 & 1579 \\
			ftv100 & 101 & 1788  & 1791 & 5974 & 1788 & 1788 \\
			\hline
			rbg323 & 323 & 1326  & 1326 & 1365 & 1328 & 1328 \\
			rbg358 & 358 & 1163  & 1163 & 1180 & 1165 & 1163 \\
			rbg403 & 403 & 2465  & 2465 & 2473 & 2465 & 2465 \\
			rbg443 & 443 & 2720  & 2720 & 2760 & 2720 & 2720 \\
			\hline
		\end{tabular}
		\caption{For each selected TSPLIB instance, the table compares the best-tour cost obtained by our method with those returned by LKH-3 under two parameter settings, namely one run with 100 trials and one run with 1000 trials, and by TGCA. TGCA is implemented directly from the pseudocode reported in \cite{zhang2026solving}, without any additional post-processing.}
	\end{table*}

	The results are summarized in Table~\ref{tab:atsp_results}. LKH reaches the best-known solution for nearly all tested instances, consistently with its role as a strong heuristic reference algorithm in this size range. The mean-field method nevertheless remains highly competitive across the full benchmark set: it recovers the best-known tour for most instances and, when it does not, its cost generally remains close to the reference value. Importantly, this behavior is not restricted to small systems. On the largest tested instances, i.e., the rbg instances with $N>400$, the mean-field method always reaches the best-known value outperforming the LKH configurations reported in the table. The comparison with TGCA further confirm the validity of our approach. TGCA returns substantially larger costs on all of these benchmarks.
	
	Again, these results do not imply that the present implementation is a fully optimized solver. Rather, they show that the proposed information-theoretic dynamics explores the cost landscape differently, while remaining based on polynomial-time operations, and can compete with state-of-the-art heuristics across system sizes, even outperforming them in selected large instances. This is particularly notable because the current implementation has not yet been optimized from a numerical perspective.
	
	

	\begin{figure}[t!]
		\centering
		\includegraphics[width=0.75\linewidth]{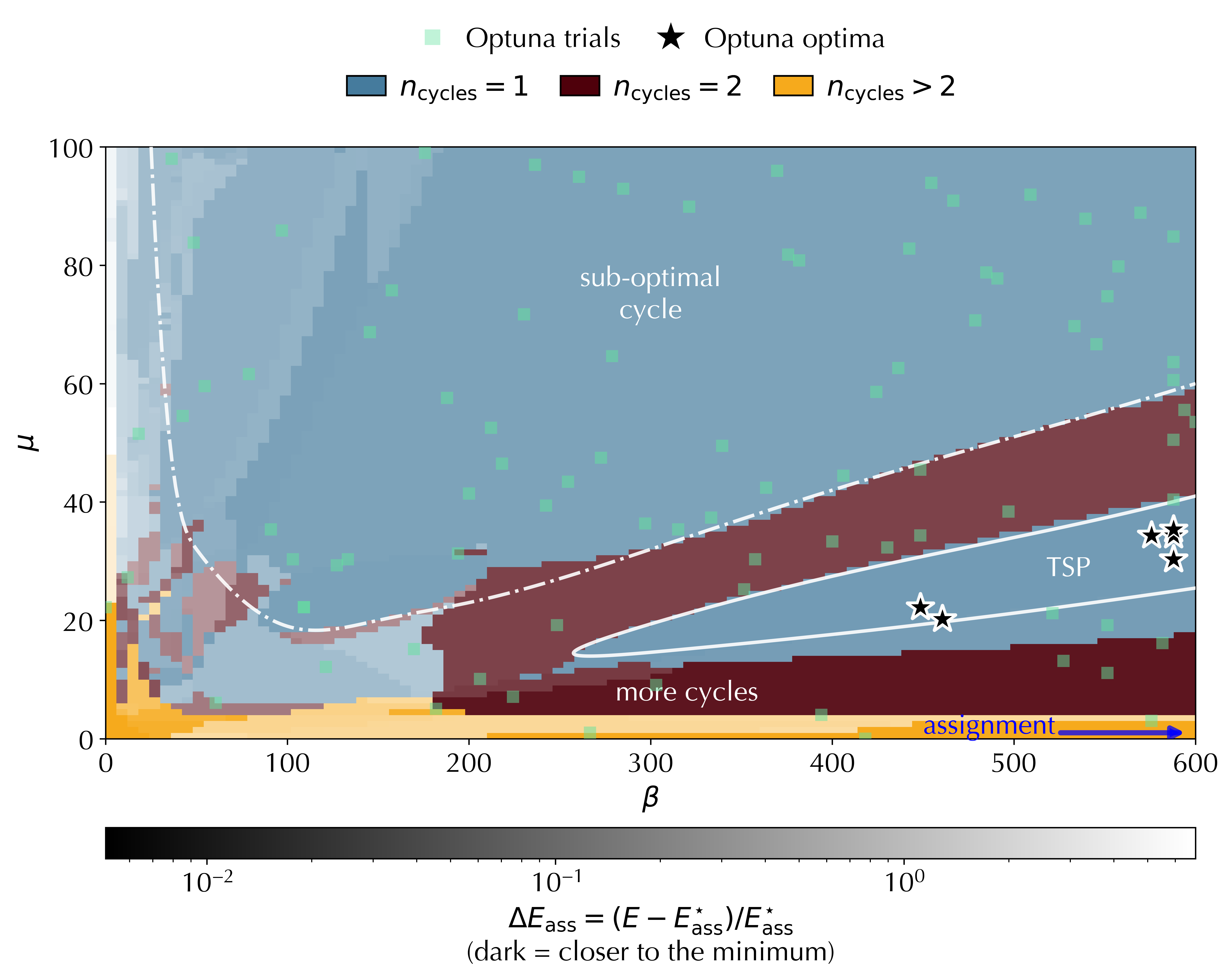}
		\caption{Diagram of the solutions obtained for a fixed cost matrix $c$ as a function of $\mu$ and $\beta$. 
			Colors indicate the number of cycles in the extracted solution, and their intensity represents the relative energy gap with respect to the assignment-problem minimum. 
			The solid white line marks the parameter region where the TSP solution is recovered, while the dashed line indicates sub-optimal connected solutions. 
			Green dots show the points sampled by Optuna, and stars mark successful trials in which the correct TSP solution is found.}
		\label{fig:diagram}
	\end{figure}
	
	To highlight the numerical challenges involved, we make one final observation. Given a cost-matrix instance, when the method does not recover the optimal TSP solution, the failure is not necessarily due to a lack of convergence of the iterative procedure. It may instead reflect the structure of the underlying solution landscape, which can depend on several features of the cost matrix: the number of subtours in the corresponding linear-assignment solution, the sizes of these subtours, frustration in the cost structure, the initial condition, and other instance-specific factors. This potentially complex landscape affects the region of the ($\beta,\mu$) parameter space in which the optimal solution can be recovered. Since both parameters must remain finite in any numerical implementation, their choice becomes part of the practical performance of the method. In the simulations presented above, we select the best combination of $\beta$ and $\mu$ using Optuna \cite{akiba2019optuna}, a hyperparameter-optimization library, as described in the Methods section.
	
	In Fig.~\ref{fig:diagram} we illustrate an example of the plane $(\beta,\mu)$ of a representative cost matrix.
	For each point of the plane, we solve the mean-field equations, extract a discrete solution, and classify it by the number of cycles and by its energy gap from the assignment-problem minimum.
	
	Although the diagram should not be interpreted as a universal phase diagram, as changing the cost matrix or the initial conditions modifies the detailed shape of the regions (see SI), it reveals a generic pattern: consistently with the theory, TSP solutions are found at large $\beta$ and nonzero $\mu$. Moreover, as $\beta$ increases, larger values of $\mu$ are also required to prevent the global penalty from becoming negligible with respect to the energy term in~\eqref{eq:lagrangian_probability}. Also, starting from the large-$\beta$ region where the linear-assignment solution is recovered, increasing $\mu$ first drives the system from the assignment-like regime toward the TSP region. However, even larger values of $\mu$ can move the solution away from this region again, producing either non-Hamiltonian configurations or connected but suboptimal tours. Thus, the relevant regime is not the limit of the strongest possible penalty, but rather a parameter region in which $\beta$ and $\mu$ are not only large enough, but also properly balanced. 

	\section{Discussion and Conclusion}
	In this paper, we introduced a framework based on information-theoretic principles for addressing  the well-known TSP. Specifically, we derive a probability distribution over graph so that, as the two key parameters $\beta$ and $\mu$ increase, it becomes increasingly concentrated on graphs corresponding to low-cost Hamiltonian cycles, that is, candidate solutions to the TSP. Building on this theoretical foundation, we developed a mean-field algorithm and benchmarked it on several TSP instances. The results show that, at least for small and medium sized instances, the proposed method is competitive with state-of-the-art heuristic algorithms.
	
	From a theoretical perspective, our approach provides an alternative way of handling the nonlinear constraints that characterize the TSP, such as subtour-elimination constraints. In particular, these constraints are embedded into a term that tilts the probability distribution over graphs toward the region corresponding to valid TSP solutions. Without this term, the probability distribution represents graphs belonging to the assignment polytope, where solutions containing subtours are allowed. Introducing the additional structural term restricts the distribution to the subset of the assignment polytope that overlaps with the much more complex TSP polytope. 
	
	From the perspective of numerical implementation, our mean-field algorithm differs fundamentally from classical statistical-physics-inspired algorithms to the TSP, such as simulated annealing, mean-field annealing, or softassign. These approaches rely on a position-based encoding, in which the optimization variables $v_{i,t}$ specify which city $i$ is assigned to which position $t$ in the tour. Under this encoding, the analogue of Eq.~(\ref{eq:mf_self_consistency}) contains a quadratic cost term, since the travel cost couples variables associated with consecutive tour positions, through contributions of the form $v_{i,t}v_{j,t+1}$. The cyclic structure of the tour is therefore built directly into the encoding, and the annealing dynamics is thus controlled only by $\beta$. By contrast, in our formulation the variables $V_{ij}$ are directly associated with the edges of the graph, as is more common in the operations-research literature \cite{applegate2011traveling}. Hence, the travel-cost contribution remains linear, while the cyclic structure is controlled separately by $\mu$. Our method therefore evolves over a two-parameter space.
	
	The superior performance of our algorithm relative to standard statistical-physics-inspired approaches (see SI) suggests that the edge-based encoding provides a more favorable optimization landscape and that it is advantageous to place the main nonlinearity in the constraint term rather than in the travel-cost contribution.
	
	Understanding the mechanisms underlying this behavior will require a deeper theoretical analysis of the phase space and of how the relative strengths of the two parameters, $\beta$ and $\mu$, affect the performance of the algorithm. In Fig.\ref{fig:diagram}, we provide only an illustrative example of the complexity of this phase space. Future studies should investigate whether general patterns emerge, including the possible presence of phase-transition phenomena.
	
	Such theoretical investigations could also guide the design of more efficient algorithms capable of scaling to larger problem sizes. In the implementation considered in this paper, the algorithm relies on a relatively simple iterative scheme with fixed values of the temperature and penalty parameters. Since we have not yet characterized the global structure of the corresponding parameter space, we did not design principled annealing schedules that could potentially improve convergence speed and computational efficiency. Nevertheless, the strong performance achieved even in this basic setting is noteworthy.
	
	Overall, our numerical results provide a foundation for the development of even more efficient TSP meta-heuristic. More importantly, this paper introduces an information-theoretic framework that extends beyond the TSP and may be generalized to a broad class of combinatorial optimization problems.


	.
	
	\section{Materials and Methods}
	
	\subsection{Pseudo-code}
	
	\begin{center}
		\begin{tcolorbox}[
			enhanced,
			width=0.96\linewidth,
			colback=gray!4,
			colframe=black!65,
			boxrule=0.6pt,
			arc=2mm,
			left=2mm,
			right=2mm,
			top=1mm,
			bottom=1mm,
			title={\textbf{Main Algorithm.}},
			coltitle=black,
			colbacktitle=gray!18,
			fonttitle=\small,
			attach boxed title to top left={xshift=2mm,yshift=-2mm},
			boxed title style={
				colback=gray!18,
				colframe=black!65,
				boxrule=0.5pt,
				arc=1.5mm
			}
			]
			
			\begin{algorithmic}[1]
				
				\Require Cost matrix $C \in \mathbb{R}^{N \times N}$, inverse temperature $\beta$, cycle enforcing parameter $\mu$, maximum iterations $T_{\max}$, tolerance $\varepsilon$
				\Ensure A candidate tour and its cost
				
				\State Initialize a soft assignment matrix $V^{(0)}$ with $V_{ij}^{(0)} \geq 0$
				
				\For{$t = 0,1,\dots,T_{\max}-1$}
				
				\State Compute the cycle enforcing matrix (with $K=N-1$):
				\[
				\Lambda^{(t)}
				=
				\sum_{k=2}^{K}
				\alpha_k
				\left((V^{(t)})^{k-1}\right)^\top .
				\]
				
				\State For fixed $\Lambda^{(t)}$, determine $u$ and $v$ from the nonlinear constraint equations (solved iteratively by fixed-point correction until the row and column residuals are below tolerance)
				\[
				\sum_j 
				\frac{1}
				{1+\exp\left(
					\beta c_{ij}-u_i-v_j+\mu\Lambda^{(t)}_{ij}
					\right)}
				=1
				\qquad \forall i,
				\]
				\[
				\sum_i 
				\frac{1}
				{1+\exp\left(
					\beta c_{ij}-u_i-v_j+\mu\Lambda^{(t)}_{ij}
					\right)}
				=1
				\qquad \forall j.
				\]

				\State Update edge probabilities:
				\[
				V_{ij}^{(t+1)}
				=
				\frac{1}
				{1+\exp\left(
					\beta c_{ij}
					- u_i
					- v_j
					+ \mu \Lambda^{(t)}_{ij}
					\right)}
				\]
				
				
				
				\If{$\|V^{(t+1)} - V^{(t)}\| < \varepsilon$}
				\State \textbf{break}
				\EndIf
				
				\EndFor
				
				\State Decode a hard assignment $A^{(t)}$ by solving the maximum-weight assignment problem
				\begin{equation*}
					A^{(t)} = \arg\max_{A \in \mathcal{A}} \sum_{i,j} A_{ij} V^{(t)}_{ij},
				\end{equation*}
				where $\mathcal{A}$ is the set of permutation matrices.
				\If{$A^{(t)}$ induces a single Hamiltonian cycle}
				\State Compute the tour cost
				\[
				E=\sum_{i,j} A^{(t)}_{ij}c_{ij}
				\]
				\State \Return candidate tour and cost $E$
				\Else
				\State The algorithm fails to produce a feasible TSP tour
				\State \Return assignment with multiple subtours
				\EndIf
				
			\end{algorithmic}
		\end{tcolorbox}
	\end{center}
	
	The dominant computational cost of the algorithm comes from the evaluation of the cycle-enforcing matrix $\Lambda^{(t)}$. Computing a single dense matrix power requires $O(N^3)$ operations, and the summation over $k=2,\dots,N-1$ therefore requires $O(N^4)$ operations per outer iteration. The solution of the nonlinear equations for the Lagrange multipliers $u$ and $v$ requires repeated evaluations of the dense matrix $V$, each costing $O(N^2)$. Therefore, for $T_{\max}$ outer iterations, the worst-case complexity is dominated by the cycle-enforcing term and scales as $O(T_{\max}N^4)$.
	
	In the simulations reported above we replace steps 4 and 5 by enforcing the assignment constraints through a Sinkhorn-type approximation \cite{sinkhorn1967concerning}, as we observe empirically that these modifications make the algorithm faster and more stable. In practice, instead of solving explicitly for the Lagrange multipliers $u$ and $v$, we form
	\[
	W_{ij}=\exp\left(-\beta c_{ij}-\mu\Lambda^{(t)}_{ij}\right)
	\]
	and alternately rescale rows and columns to obtain
	\[
	V^{(t+1)}=\mathrm{diag}(a)\,W\,\mathrm{diag}(b),
	\qquad
	V^{(t+1)}\mathbf{1}=\mathbf{1},
	\qquad
	V^{(t+1)\top}\mathbf{1}=\mathbf{1}.
	\]
	For additional numerical stability, we apply damping and set $V^{(t+1)}=(1-d)V^{(t)}+d V^{(t+1)}$ where $d\in(0,1]$ is the damping parameter. 
	
	For larger instances (especially for the largest matrices in the benchmark table), we implement a second approximation to reduce the computational cost. In step 3, instead of computing the cycle-enforcing term from the dense matrix $V$, we use the binary permutation matrix obtained by solving the assignment problem with weights $V$, following the approach of \cite{zhang2026solving}. Since permutation matrices are sparse and binary, their powers correspond to compositions of the same permutation, reducing the computational cost of the cycle-enforcing term from $O(N^3)$ to $O(N^2)$. 
	
	Regarding the choice of parameters, we first set $\alpha_k=1$ for any $k$. Then, we know that for $\beta \to \infty$ and sufficiently large $\mu$, the matrix $V$ is expected to converge to a binary matrix representing a low-cost Hamiltonian tour (when the optimal solution is not unique, $V$ instead converges to a probabilistic representation of the links appearing across the degenerate solutions). In practice, however, $\beta$ is finite, and identifying the best combination of $\beta$ and $\mu$ is nontrivial.
	We therefore tune the parameters using Optuna \cite{akiba2019optuna}, a hyperparameter-optimization framework that efficiently explores the parameter space. The objective function is the tour cost, with a large penalty assigned whenever the algorithm fails to produce a single Hamiltonian cycle. Specifically, Optuna searches over $\beta$, $\mu$, the damping parameter $d$, and the maximum cycle length $K$ used in the cycle-enforcing term (step 3). The latter is also treated as a tunable parameter because, in many instances, penalizing all possible cycle lengths is not necessary to obtain a single-cycle assignment, and reducing $K$ lowers the computational cost.
	
	Finally, for the initial condition $V^{(0)}$, we generally use a random non-negative matrix. However, we empirically observed that for some benchmark instances, especially the rbg benchmark matrices, the optimization is more efficient when the initial condition is biased toward low-cost edges. In these cases, we initialize
	$V^{(0)}_{ij} \propto \exp(-c_{ij})$
	so that edges with lower cost have larger initial weights.
	
	\bibliographystyle{unsrt}
	\bibliography{bib}
	
\end{document}